# Video-rate broadband longwave IR dual-comb spectroscopy with 240,000 comb-mode resolved data points

SERGEY VASILYEV[1,*], ANDREY MURAVIEV[2], DMITRII KONNOV[2], MIKE MIROV[1], VIKTOR SMOSLKI[1], IGOR MOSKALEV[1], SERGEY MIROV[3], AND KONSTANTIN VODOPYANOV[2]

[1]*IPG Photonics – Southeast Technology Center, Birmingham, AL 35211, USA*
[2]*CREOL, College of Optics and Photonics, University of Central Florida, Orlando, Florida 32816, USA*
[3]*Department of Physics, the University of Alabama at Birmingham, Birmingham, AL 35294, USA*

We report molecular detection using dual frequency-comb spectroscopy with highly coherent broadband (6.6–11.4 μm) long-wavelength infrared (LWIR) combs. The combs were produced via intra-pulse difference frequency generation (IDFG) in ZGP crystals using sub-three-cycle (20 fs) driving pulses from mode-locked Cr:ZnS lasers at the central wavelength λ=2.4 μm. Real-time and up to video rate (0.1–12 s per spectrum) acquisition of molecular spectra with some 240,000 comb-mode-resolved data points spaced by 80 MHz and referenced to a Rb clock with a signal-to-noise ratio (SNR) >300 has been demonstrated. The key to achieving such a high rate of massive spectral data acquisition is the low phase and intensity noise of the LWIR combs and excellent mutual coherence. The high SNR was also facilitated by the high (7.5%) IDFG conversion efficiency resulting in an average LWIR comb power of 300 mW per channel.

## 1. Introduction

Dual-comb spectroscopy (DCS), a subtype of Fourier-transform spectroscopy, organically exploits high spatial and temporal coherence of optical frequency combs. DCS in the mid-infrared 'functional group' region across the 3–6 μm region [1, 2, 3] and long-wavelength infrared (LWIR) 'molecular fingerprint' (6.7–20 μm) range of the spectrum [4, 5] are desirable from the viewpoint of high-sensitivity detection. Several LWIR DCS systems have been reported with broadband (optical octave or similar) combs based on down-conversion of well-developed mode-locked fiber lasers. Timmers et al. presented a robust method for generating LWIR combs through intra-pulse difference frequency generation (IDFG) driven by spectrally broadened and compressed few-cycle pulses with a 10.6-fs duration from an Er-doped fiber laser system [4]. Because IDFG modes occur as a difference frequency between the modes of the same pump laser comb, the carrier-envelope offset ($f_{ceo}$) frequency of the pump is cancelled out, providing an offset-free comb at the laser repetition frequency ($f_{rep}$). With the pump power of 350 mW the authors produced, using an orientation-patterned GaP crystal, a comb spanning 4–12 μm with 0.25 mW average power and demonstrated comb-mode resolved (spacing 100 MHz) DCS of methanol and ethanol at 50 mbar pressure with the averaging time 500 s. By reducing the interferogram window from 20 ms to 285 μs and dropping the spectral resolution to 7 GHz, the authors managed to reduce the averaging time to 2 s. [4]. Kowligy et al. used a new approach to DCS that unites the IDFG method for generating a frequency comb with electro-optic (EO) detection in GaSe crystal; this approach expands the dynamic range of detection and shows promise for fast spectral measurements [5]. Although the system made it possible to obtain spectra with the comb resolution (100 MHz), spectroscopic measurements were carried out with a broadband absorber ($CO_2$ band at λ≈15 μm; 1 atm pressure; linewidth 5 GHz), and the authors used apodization so that the resolution dropped to 500 MHz [5]. On the other hand, frequency combs based on quantum cascade lasers (QCLs) are well suited for performing DCS with high (few μs) time resolution, thanks to their high (~10 GHz) repetition rate. By shifting QCL combs via current modulation, the sampling density can be improved to few MHz by interleaving multiple spectra [6, 7]. However, the frequency axis of QCL combs must be calibrated against a known absorption line for each measurement, and they have a relatively narrow (50–60 $cm^{-1}$) spectral bandwidth per device. Thus, the existing LWIR DCS methods provide fast spectral measurements either with broadband coverage but low resolution or high resolution but narrow spectral coverage.

Here we present a new platform for performing broadband (span 625 $cm^{-1}$) comb-mode-resolved (80 MHz) LWIR DCS with acquisition rates up to video frequency (10 Hz) that allows obtaining metrological-quality spectra in seconds. The platform is based on the IDFG with a pair of robust, low-noise Cr:ZnS laser frequency combs operating at 2.4 μm.

## 2. Cr:ZnS-ZGP frequency comb

The design of the Cr:ZnS frequency comb source, the structure of its spectrum, and the comb referencing scheme are illustrated in Fig 1. The key components of the Cr:ZnS comb are: a high average power (1.4 W) 3-optical-cycle (24 fs) polycrystalline Cr:ZnS master oscillator (MO) at the central wavelength of 2.4 µm and repetition rate $f_{rep} \approx 80$ MHz [8] and a polycrystalline Cr:ZnS power amplifier/supercontinuum generator (PA-SCG) [9]. The MO and the PA-SCG are optically pumped by a continuous-wave single frequency (SF) source consisting of a semiconductor seed laser at 1567 nm and two commercial Er-doped fiber amplifiers that are optimized for the SF regime. Typical pump power levels of the MO and PA-SCG stages are 7 W and 10 W, respectively.

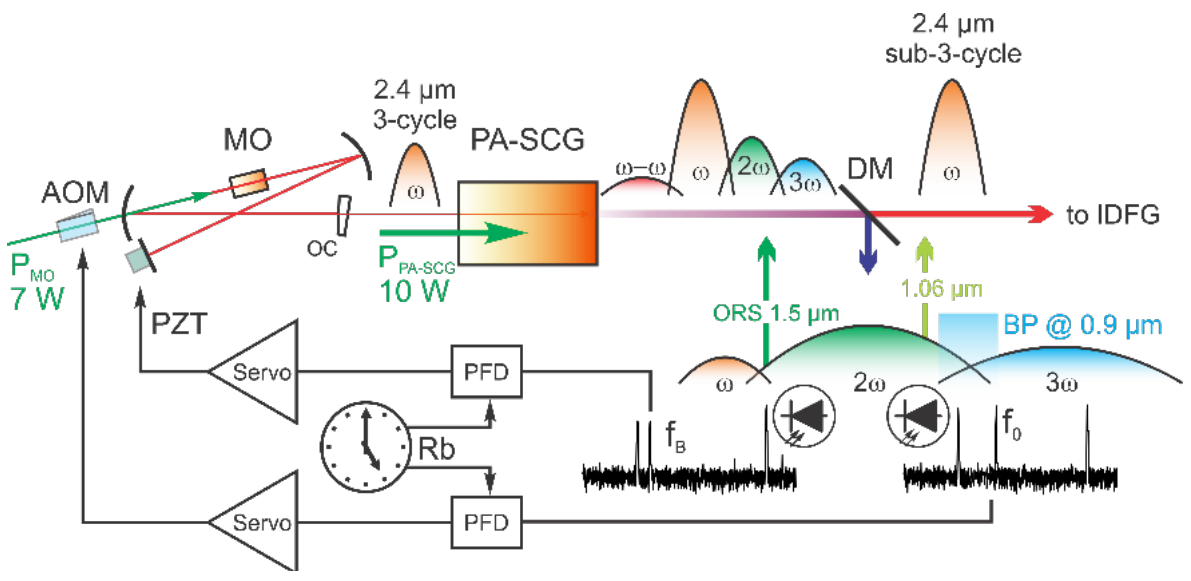

Fig. 1. Schematic of the Cr:ZnS frequency comb setup. MO, Master oscillator; PA-SCG, power amplifier/supercontinuum generator; $P_{MO}$ and $P_{PA\text{-}SCG}$, are optical pumps for MO and PA-SCG, respectively; ω-ω, ω, 2ω, 3ω, frequency combs that comprise the supercontinuum coming from the laser. DM, dichroic mirror; BP, optical bandpass filter; AOM, acousto-optic modulator; PZT, piezo transducer; PFD, phase frequency detector; Rb, synthezed RF signal referenced to a Rb clock; ORS, narrowband laser for optical referencing.

During propagation through the PA-SCG stage, the MO seed pulses are simultaneously amplified to about 4 W, their spectrum is broadened to an optical octave, and the pulse width is compressed to about 20 fs. Further, the random quasi-phase matching (RQPM) process in polycrystalline Cr:ZnS results in a series of intrapulse three-wave mixings [10]. Thus, the spectrum at the output of the PA-SCG stage consists of several superimposed frequency combs corresponding to different mixing processes: ω-ω=0 (intra-pulse difference frequency generation); ω+ω=2ω; ω+2ω=3ω (2[nd] and the 3[rd] harmonic generation), etc., with carrier-envelope offset (CEO) frequencies correspondingly 0, $f_0$, $2f_0$, $3f_0$ and the mode spacing ($f_{rep}$) defined by that of the oscillator. Here ω is the center angular frequency of the Cr:ZnS oscillator and $f_0$ is its CEO frequency.

The near-IR part of the continuum is separated from the fundamental comb at 2.4 µm by a dichroic mirror (DM) and used for the comb's optical referencing. The offset frequency $f_0$ is measured at the wavelength 0.9 µm with a Si avalanche photodetector via an interference beating between the 2ω and 3ω combs (2$f$-3$f$ interferometry). Further, spectral components of the continuum near the wavelength 1.54 µm (the 2ω comb) are used for the combs' optical referencing to an ORS-mini ultra-stable laser provided by Menlo Systems: a relevant part of the 2ω spectrum is bandpass filtered and superimposed with the reference laser on a 50:50 fiber optic coupler and then an interference beating $f_B$ between the reference laser and spectral components of the continuum are detected with a balanced InGaAs photodetector. The measured $f_0$ and $f_B$ signals are phase-locked to synthesized radiofrequency (RF) signals that, in turn, are referenced to a Rb clock. The $f_0$ signal is controlled by the modulation of the oscillator pump power with an acousto-optic modulator (AOM) and the $f_B$ signal is controlled by the adjustment of the oscillator cavity length with a piezo transducer (PZT), as described in detail in [11].

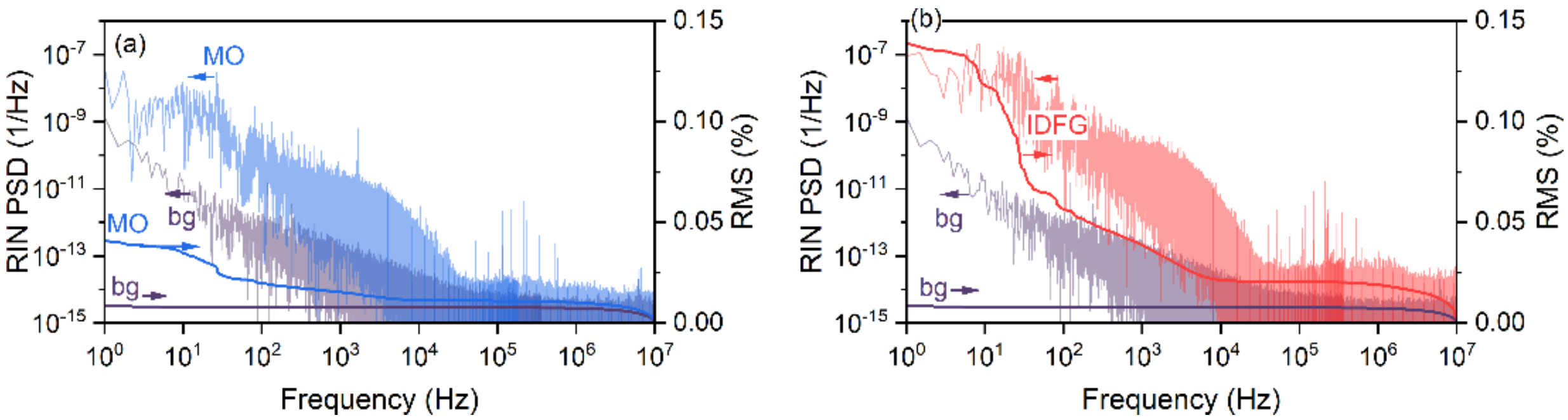

Fig. 2. (a) Intensity noise of the Cr:ZnS oscillator at 2.4 µm. (b) Intensity noise of the IDFG comb. Left axes: RIN power spectral density (PSD); right axes: RMS intensity noise.

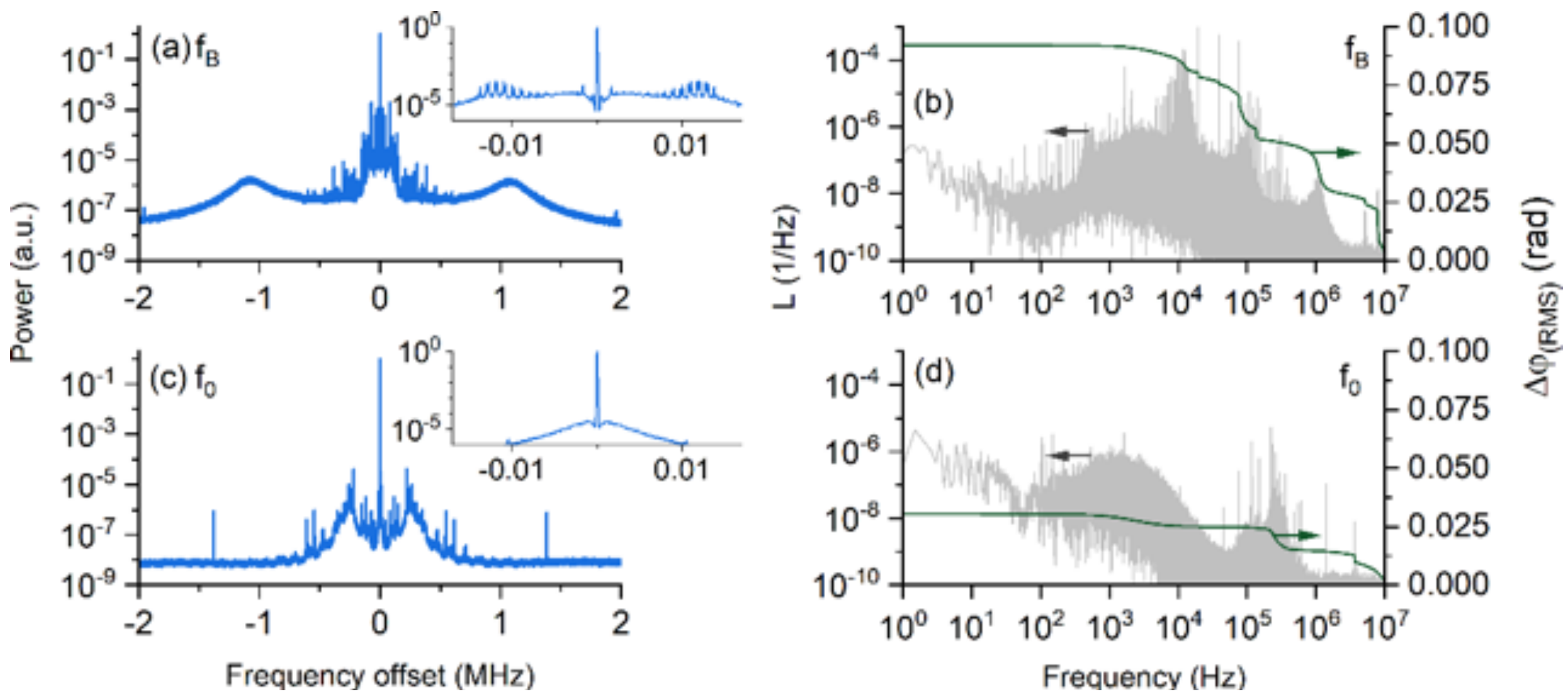

Fig. 3 Parameters of the phase-locked $f_B$ and $f_0$ signals measured in loop. (a, c) power spectra of $f_B$ and $f_0$ signals, respectively (100-Hz resolution bandwidth). The insets show coherence peaks in the central parts of the spectra. (b, d) phase noise of $f_B$ and $f_0$ signals. Left axes, gray color: phase noise (PSD). Right axes, solid black curves: RMS phase noise.

## 3. Intra-pulse difference frequency generation and noise measurements

To produce a LWIR comb, the main output at 2.4 µm was focused into a 3-mm-long AR-coated $ZnGeP_2$ (ZGP) crystal (type I phase matching with polar and azimuthal angles $\theta \approx 51°$ and $\phi \approx 0°$, correspondingly [12]) using an $f$=50-mm 90º off-axis parabolic mirror. Typically, when propagating through the ZGP crystal, the high peak power of few-cycle laser pulses at 2.4 µm leads not only to the formation of an IDFG comb spanning approximately 5–12 µm, but also to the broadening of the laser spectrum and its redshift. The redshift is caused by pumping 'blue' photons into the 'red' part of the spectrum due to the IDFG process (it is also dictated by energy conservation). As a result, the broadened spectrum of the laser (with $f_{ceo}= f_0$) and the offset-free IDFG output merge into a continuum that spans from below 1.9 µm to beyond 12 µm [13,14]. In our dual-comb spectroscopy experiments, the IDFG comb was separated by using a longpass (>6.7 µm; -3dB) filter; its average power was measured to be as high as 300 mW.

The relative intensity noise (RIN) of the Cr:ZnS oscillator and of the IDFG comb were measured using a fast DC-coupled HgCdTe photodetector (77K, 50 MHz, Kolmar Technologies) and fed into a 16-bit, 250 MHz sampling rate A/D converter (AlazarTech ATS9626). The photodetector DC offset and the optical signal level at its input were adjusted to access the whole dynamic range of the digitizer. The digitized signal (1 Gigasample at the sampling interval of 4 ns) was post-processed on a computer.

The results are summarized in Fig. 2. The RIN of the oscillator integrated from 1 Hz to 10 MHz was 0.032% RMS, which is below the levels achieved with other purpose-build low-noise mode-locked Cr:ZnS laser oscillators [14,15, 16]. We attribute this property to the high second harmonic generation (SHG) level in the oscillator's polycrystalline gain medium (5–7% of the main signal). The high instantaneous SHG losses result in a noise suppression, as described in [17]. Conversely, our measurements show that the IDFG process increases the intensity noise by about a factor of 4 (Fig. 2b).

The phase noise of the phase-locked $f_0$ and $f_B$ signals was measured in-loop by decoupling 10% of the RF signals' power. Decoupled signals were properly low-pass filtered, amplified, digitized and then post-processed. The obtained results are summarized in Fig. 3. The signal-to-noise ratio of the coherence peaks was 50 dB in a 100-Hz resolution bandwidth in both cases. The integrated (1 Hz to 10 MHz) phase noise of the $f_B$ and $f_0$ signals was 0.09 rad and 0.03 rad, respectively, which corresponds to robust phase locking. Additionally, we performed the analysis of the heterodyne beats $f_B$ for a free-running Cr:ZnS oscillator. The estimated linewidth of the free running comb was as low as 12.5 kHz in a 0.2-ms window (see Supplement 1).

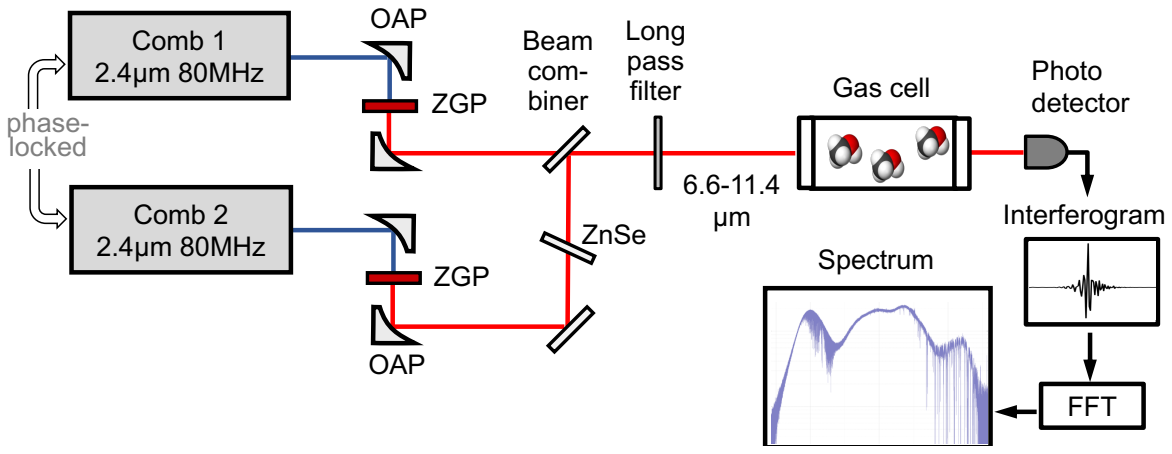

Fig. 4 Experimental layout for the LWIR dual-comb spectroscopy. OAP, off-axis parabolic mirror; ZnSe, dispersion compensator plate.

## 4. Dual-comb spectroscopy experiment

The DCS experiment was carried out by using two identical mutually phase-locked offset-free IDFG combs operating at repetition rates near 80 MHz with the effective bandwidth (-20 dB level) of 880–1505 $cm^{-1}$ (6.6–11.4 μm). On the short-wave side, the comb spectrum was limited by the longpass filter, and on the long-wave side – by the HgCdTe photodetector cutoff. The two IDFG combs were superimposed with a broadband beam combiner (Fig. 4) and sent through a gas cell (a *symmetric* DCS scheme). The periodic beatings (interferograms) between the combs at $\Delta f_{rep} = f_{rep2} - f_{rep1}$ =131 Hz were detected with the HgCdTe photodetector, digitized, and processed on a PC. To avoid the folding of the RF spectrum, the digitizer input was lowpass filtered to <40 MHz.

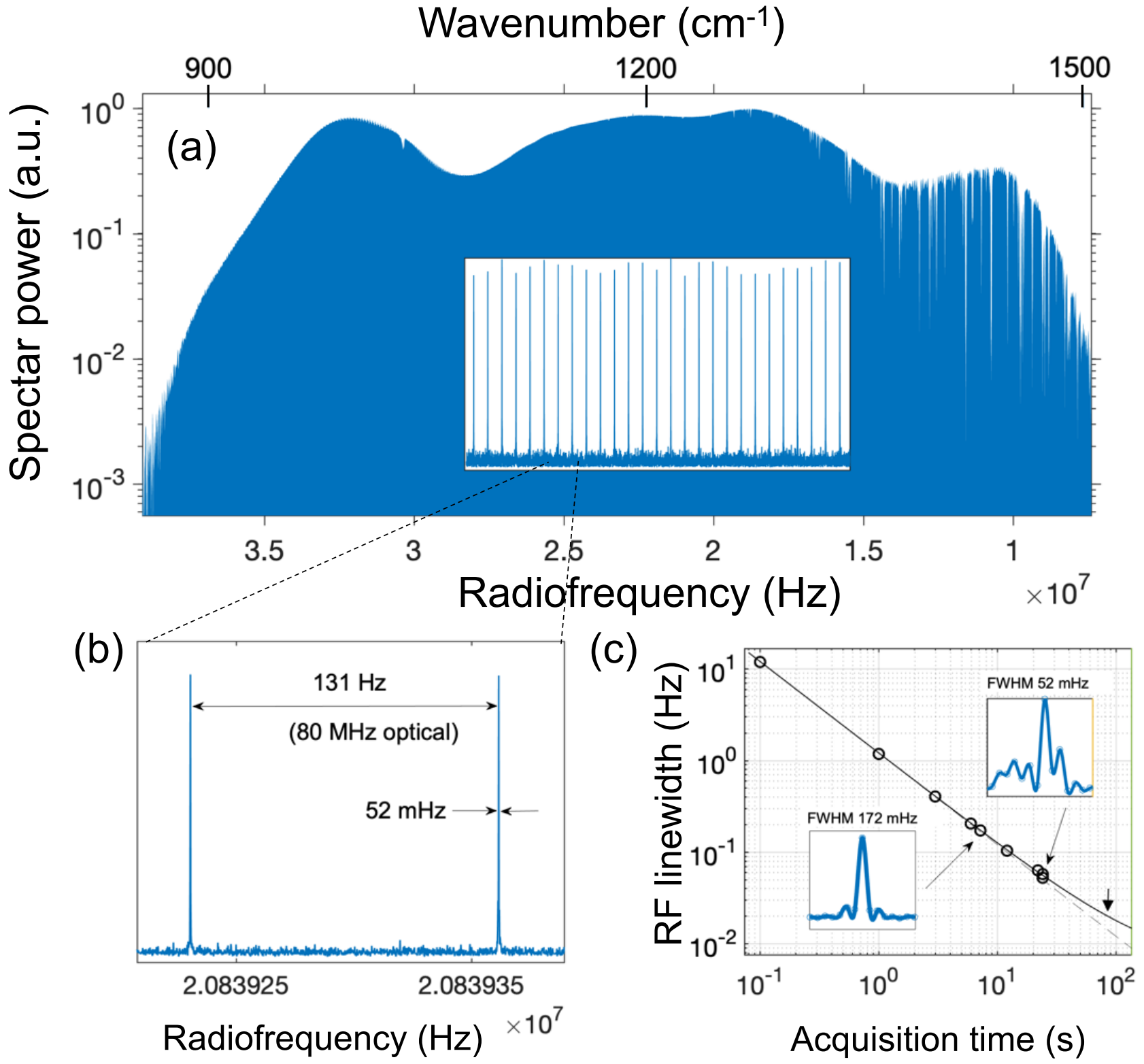

Fig. 5 (a) Mode-resolved spectrum of the LWIR comb. The bottom is the RF and the top is the optical frequency scale. Spectral dips between 1300 and 1500 cm-1 are due to water absorption in the beam path. (b) Zoomed-in RF spectrum that shows comb lines with the finesse of >2,500. (c) RF comb FWHM linewidth as a function of the acquisition time for the stream of multiple interferograms. The dashed line corresponds to the time-bandwidth limited case. Solid line: best fit to our data. The vertical arrow indicates the mutual coherence time of 85 s.

To obtain the mode-resolved DCS spectra, we recorded streams of data containing multiple sequential interferograms spaced by $1/\Delta f_{rep}$ = 7.6 ms, with a time window varying from 0.1 to 24 s and no apodization. A single interferogram's signal-to-noise ratio (SNR) was measured to be SNR>400. Fig. 5a shows a mode resolved RF comb spectrum, obtained via Fast Fourier Transform (FFT) of the signal from the photodetector. The optical frequency scale (shown on top) was retrieved from the $f_{rep}$ and $\Delta f_{rep}$ obtained from the readings of frequency counters referenced to a Rb clock. The RF scale is inverted since the beats occurred between the $n$-th mode of one comb and the $(n+1)$-th mode of the other. The zoomed RF spectrum at 24-s time window (Fig. 5b) reveals a comb-mode structure with 131-Hz spacing (80 MHz optical) and the linewidth of 52 mHz, corresponding to the finesse of >2,500. Fig. 5c plots a log-log dependence of the full-width-half-maximum (FWHM) RF linewidth vs. the recording time window (τ), which shows that for τ <10 s it follows the time-bandwidth limit behavior (FWHM linewidth ≈1.2/τ) represented by the dashed line. At longer τ, we observe a deviation from this law with the solid curve being the best fit. By extrapolating this curve, we get the result that the line broadens to $\sqrt{2}$ times of its time-bandwidth limit at τ≈85 s (a vertical arrow in Fig. 5c indicates this), which we consider being the mutual coherence time between the two LWIR combs.

We used two low-pressure samples to attain high-resolution molecular spectra with our DCS system: nitrous oxide ($N_2O$) and methanol ($CH_3OH$). Here we coherently averaged single DCS interferograms, each with a time window of $1/\Delta f_{rep}$ = 7.6 ms, with no apodization, and the averaging time of 12 s (1573 interferograms).

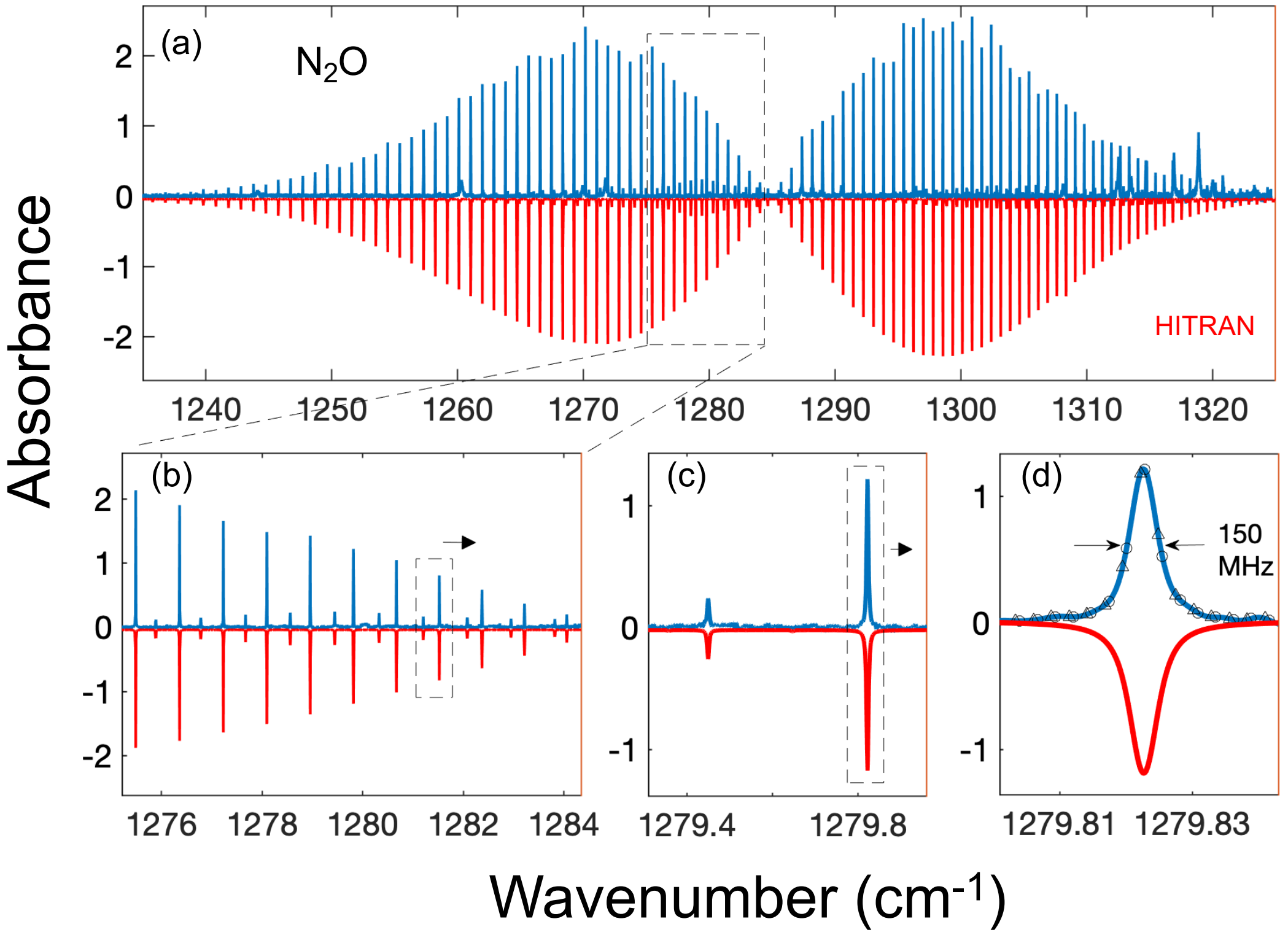

Fig. 6. (a) Absorbance spectrum of the ν1 fundamental band of N2O taken with the averaging time of 12s. (b-d) Zoomed-in portions of the spectrum. Shown in (d) as circles and triangles are spectral positions of the pairs of frequency comb modes that generate RF beats in DCS. Simulated (HITRAN) spectra are shown in red and inverted for clarity.

For nitrous oxide, we used a 12-cm-long optical cell filled with 1% $N_2O$ diluted in $N_2$ at a total pressure of 18 mbar. Figure 6 shows the absorbance spectrum $A = \ln(I_0/I)$, where $I_0$ and $I$ are the light intensities before and after the sample, respectively, corresponding to the $\nu_1$ fundamental band of $N_2O$ centered at 1285 $cm^{-1}$ (7.78 μm). The spectrum was obtained by subtracting the baseline, approximated with a polynomial fit. Also shown is the simulated spectrum (inverted in sign) from the HITRAN2020 database [18]. The peaks at around 1260, 1272, and 1308-1320 $cm^{-1}$ in the experimental spectrum are caused by water absorption in the atmosphere. The sampling interval of 80 MHz is close to be adequate for resolving the fine structure of $N_2O$ spectrum at 18 mbar (Fig. 6d): we observed a good fit to the HITRAN simulation of peak positions, linewidths (150 MHz), and relative peak intensities. The circles and triangles in Fig. 6d indicate the positions of the pairs of comb teeth that sample the absorption and produce RF beats. In total, we resolved about 240,000 spectral data points (pairs of comb teeth) in the spectral range 880-1505 $cm^{-1}$ with SNR=312 for the strongest $N_2O$ lines.

We used a 10-cm-long optical cell filled with $CH_3OH$ vapor at approximately 2.8 mbar pressure to measure the methanol spectrum. Figure 7 shows the absorbance spectrum corresponding to the $\nu_8$ (C-O stretch) band of methanol centered at 1030 $cm^{-1}$ (9.7 μm). The spectrum agrees well with the HITRAN simulation in terms of peak positions, linewidths (108 MHz), and relative peak intensities. The circles and triangles in Fig. 7e represent the pairs of comb teeth that sample the absorption line. However, to better characterize the lineshapes (which is beyond the scope of this article), it is necessary to switch to asymmetric DCS mode and increase the density of data points by combs' shifting and interleaving several spectra [3].

To demonstrate the high resolution and high speed of molecular detection simultaneously, we varied the acquisition time from 12s (coherent averaging of 1573 interferograms) to 0.1s (13 interferograms). Fig. 8 shows a portion of the $N_2O$ absorption spectrum for different averaging times. Even at 0.1-s averaging (10 Hz rate), one can detect the peaks of $N_2O$ with SNR=23 at full spectral resolution (80-MHz sampling interval).

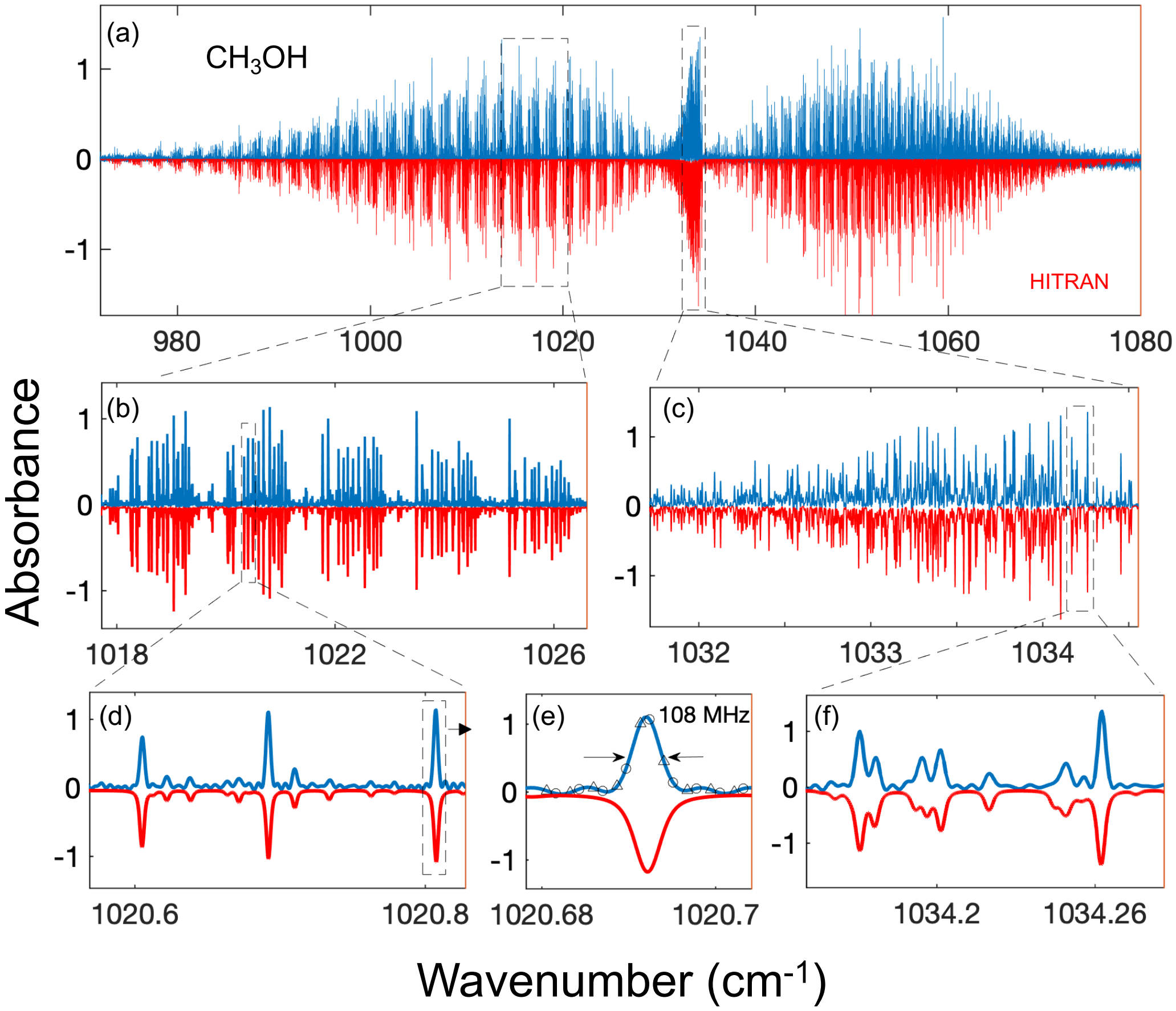

Fig. 7. (a) Absorbance spectrum of the v8 (C-O stretch) band of methanol (CH3OH) taken with the acquisition time of 12s. (b-f) Zoomed-in portions of the spectrum. Shown in (e) as circles and triangles are spectral positions of the pairs of frequency comb modes that generate RF beats in DCS. Simulated (HITRAN) spectra are shown in red and inverted for clarity.

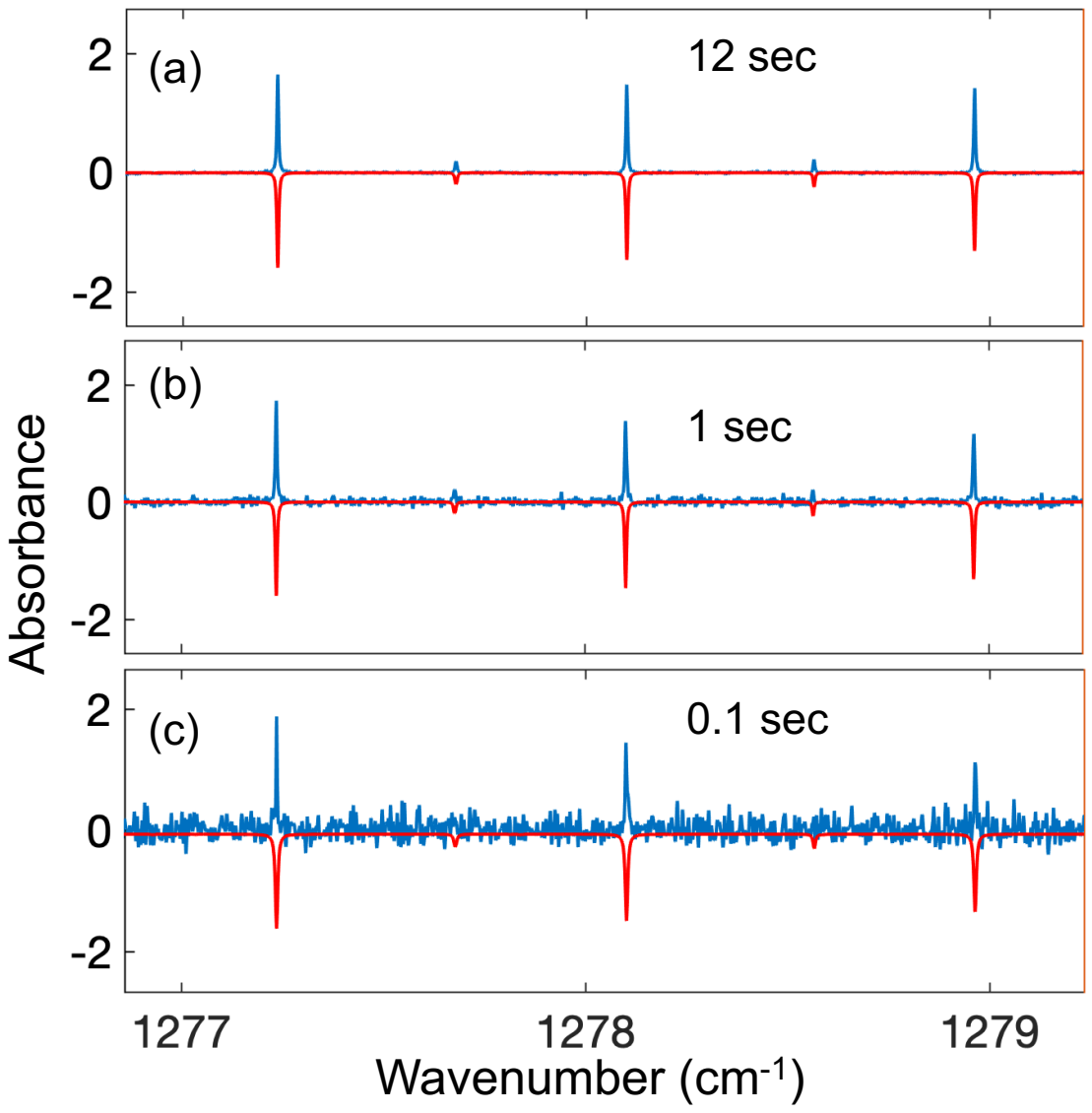

Fig. 8. (a-c) A portion of the $N_2O$ absorption spectrum taken at different averaging times (0.1-12s). Simulated (HITRAN) spectra are shown in red and inverted for clarity.

By characterizing the noise in the absorption-free parts of the spectrum, we estimate the signal-to-noise ratio of SNR=36 $Hz^{1/2}$ (here SNR=1/σ, where σ is the fractional root-mean-square deviation of the comb-mode power that scales as $1/\sqrt{\tau}$). For the M≈200,000 modes within the central (-6 dB level) 540 $cm^{-1}$-wide portion of our LWIR spectrum, this yields a DCS quality factor [19] of M×SNR = $7.3\times10^6$ $Hz^{1/2}$, a factor of 5 improvement in comparison with the quality factor for the fingerprint region reported in [4].

In summary, we performed high-resolution dual-comb LWIR spectroscopy using a pair of highly coherent (mutual coherence time 85 sec) broadband frequency combs produced by down-conversion of mode-locked Cr:ZnS lasers via IDFG in ZGP crystals. The system allowed spectral sampling with a comb-mode spacing of 80 MHz in the wavelength range of 6.6–11.4 µm and recorded 240,000 comb-mode pairs' amplitudes with up to video rate (10 Hz) and SNR=36 $Hz^{1/2}$. We have demonstrated that metrological quality LWIR absorption spectra with the frequency axis referenced to a Rb clock can be obtained in just 12 seconds. Our experiments demonstrate several advantages of ultrafast Cr:ZnS lasers at λ=2.4 μm in the context of mid-infrared DCS: (i) intrinsically low noise of the comb produced in a polycrystalline Cr:ZnS Kerr-lens mode-locked oscillator, (ii) straightforward detection and locking of $f_{ceo}$ via intrinsic nonlinear 2*f*-3*f* interferometry, and (iii) record high 7.5% IDFG efficiency in ZGP allowing to achieve 300 mW of LWIR output corresponding to 1.5 μW per comb mode, which will facilitate remote sensing as well as nonlinear Doppler-free spectroscopy of molecules. When using a nonlinear GaSe crystal, the IDFG wavelength range can be extended to 20 μm. Our next step is to utilize these advantages in real-life applications such as high-speed high-resolution molecular spectroscopy, generation of high-accuracy line lists for molecular databases, and biomedical applications.

**Funding**

S.M. and K.V. acknowledge support from the Defense Advanced Research Projects Agency (DARPA) (Grant No. W31P4Q-15-1-0008). S.M. acknowledges support from the National Institute of Environmental Health Sciences (P42ES027723) and U.S. Department of Energy (DESC0018378). K.V. acknowledges support from the Office of Naval Research (ONR) (Grant Nos. N00014-15-1-2659, N00014-18-1-2176, N00014-17-1-2705, and N68335-20-C-0251), and the Department of Energy (DOE) (Grant No. B&R No. KA2601020).

**Acknowledgments**

The authors express special thanks Ronald Holzwarth and Simon Kocur, both with Menlo Systems, for providing the Optical Reference System (ORS-mini). The authors also thank Vladimir Fedorov and Dmitry Martyshkin, both with the University of Alabama at Birmingham, and Yury Barnakov with IPG Photonics Southeast Technology Center for useful discussions.

**Supplemental document.**

See Supplement 1 for supporting content.

# Video-rate broadband longwave IR dual-comb spectroscopy with 240,000 comb-mode resolved data points

SERGEY VASILYEV[1,*], ANDREY MURAVIEV[2], DMITRII KONNOV[2], MIKE MIROV[1], VIKTOR SMOSLKI[1], IGOR MOSKALEV[1], SERGEY MIROV[3], AND KONSTANTIN VODOPYANOV[2]

[1]*IPG Photonics – Southeast Technology Center, Birmingham, AL 35211, USA*
[2]*CREOL, College of Optics and Photonics, University of Central Florida, Orlando, Florida 32816, USA*
[3]*Department of Physics, the University of Alabama at Birmingham, Birmingham, AL 35294, USA*

# Supplement 1

## 1. Generation of broad mid-IR continua in ZGP crystal

The setup for mid-IR continuum generation is illustrated in Figure S1. A Cr:ZnS frequency comb source consists of a 3-cycle (24 fs) polycrystalline Cr:ZnS master oscillator (MO) and power amplifier/supercontinuum generator (PA-SCG), both optically pumped at the wavelength 1567 nm by an amplified single-frequency (FS) semiconductor laser (seed). The output of the comb source – the pulse train at the central wavelength of 2.4 µm – is coupled to the anti-reflection (AR) coated ZGP crystal with a gold-coated off-axis parabola ($OAP_{IN}$, Thorlabs). Here a part of the input radiation is converted to offset-free long-wave IR (LWIR) transients via the optical rectification process, which is equivalent to intra-pulse difference frequency generation (IDFG) [1]. The broadband output radiation is collimated with another $OAP_{OUT}$ with the reflected focal length (RFL=15 mm, Thorlabs). A plane-parallel AR-coated plate made of undoped YAG crystal is installed before the $OAP_{IN}$ for dispersion control.

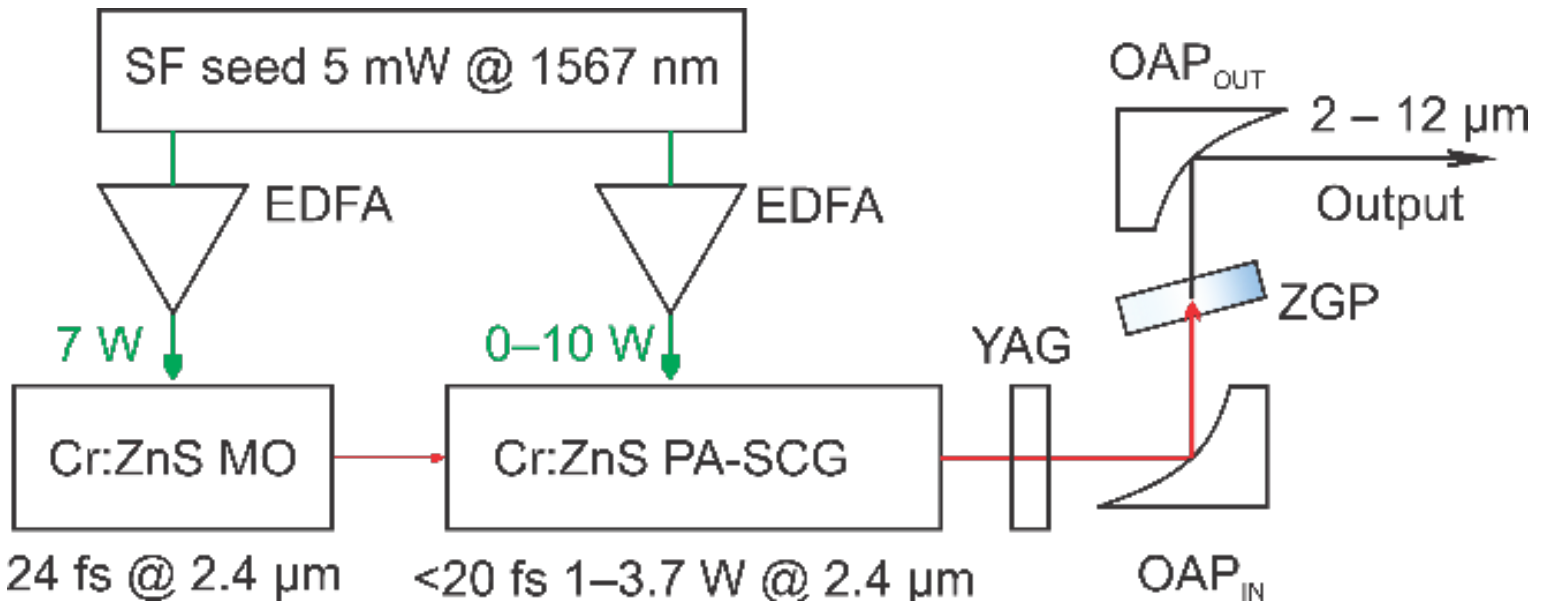

Figure S1 Mid-IR supercontinuum source (see text). EDFA, Er-doped fiber amplifier.

The thickness of the plate, ZGP crystal, and the RFL of the $OAP_{IN}$ are optimized experimentally to obtain the most uniform spectral distribution at the output of the ZGP crystal. In this particular case, we used a 2-mm thick plate and 3-mm thick ZGP sample and the $OAP_{IN}$ with the RFL=50 mm.

Figure S2 compares measured output spectra (red lines) with the spectra of input pulses from the comb source (thin blue lines). The spectra were measured at different setpoints of the comb's power amplifier. The top panel (a) corresponds to the amplifier inactive, and the bottom panel (e) corresponds to the maximum average power of 3.7 W. The spectral bandwidth and duration of amplified pulses depend on the amplifier setpoint. Further, the amplified pulses are chirped due to the amplifier's gain medium's self-phase modulation and chromatic dispersion. The experiments with similar sources [2, 3] show that the use of dispersion compensation plates allows to re-compress amplified pulses to about 120% of their Fourier transform limit (with the 20% difference arising from the uncompensated third order dispersion). Thus, we can estimate <20 fs (<2.5-optical-cycle) pulse width at the input of the ZGP crystal.

At low energies of input pulses (2 top panels in Figure S2) we observe a very low level of the offset-free LWIR signal (0f) and yet the noticeable red shift of the fundamental spectrum (f*). With an increase of energy and bandwidth of input pulses (middle panel), we observe a further red shift of the fundamental and blue shift of the 0f signal. Then at some point, we observe an explosive broadening and merging of f* and 0f spectral bands in a relatively uniform continuum. Further, we observe a significant level of nonlinear losses in ZGP crystal: transmission through the crystal drops from 83.3 % at 1.1W input power to 76.5% at 3.7 W input power. We attribute these losses to the onset of multiphoton absorption in the material [4]. Thus, the propagation of few-cycle pulses with MW peak power in ZGP crystal is quite complex because it is simultaneously governed by three-wave mixing, four-wave mixing, and self-focusing, as well as the nonlinear absorption and thermal optical effects in the bulk medium.

The average power in the different parts of the continuum was measured using Spectrogon spectral filters. The average power in the offset-free 0f band – measured behind a long-pass filter with the 6.7 µm cut-on wavelength – exceeds 0.3W, while that of the spectral band between 3 and 7 µm is about 0.5 W.

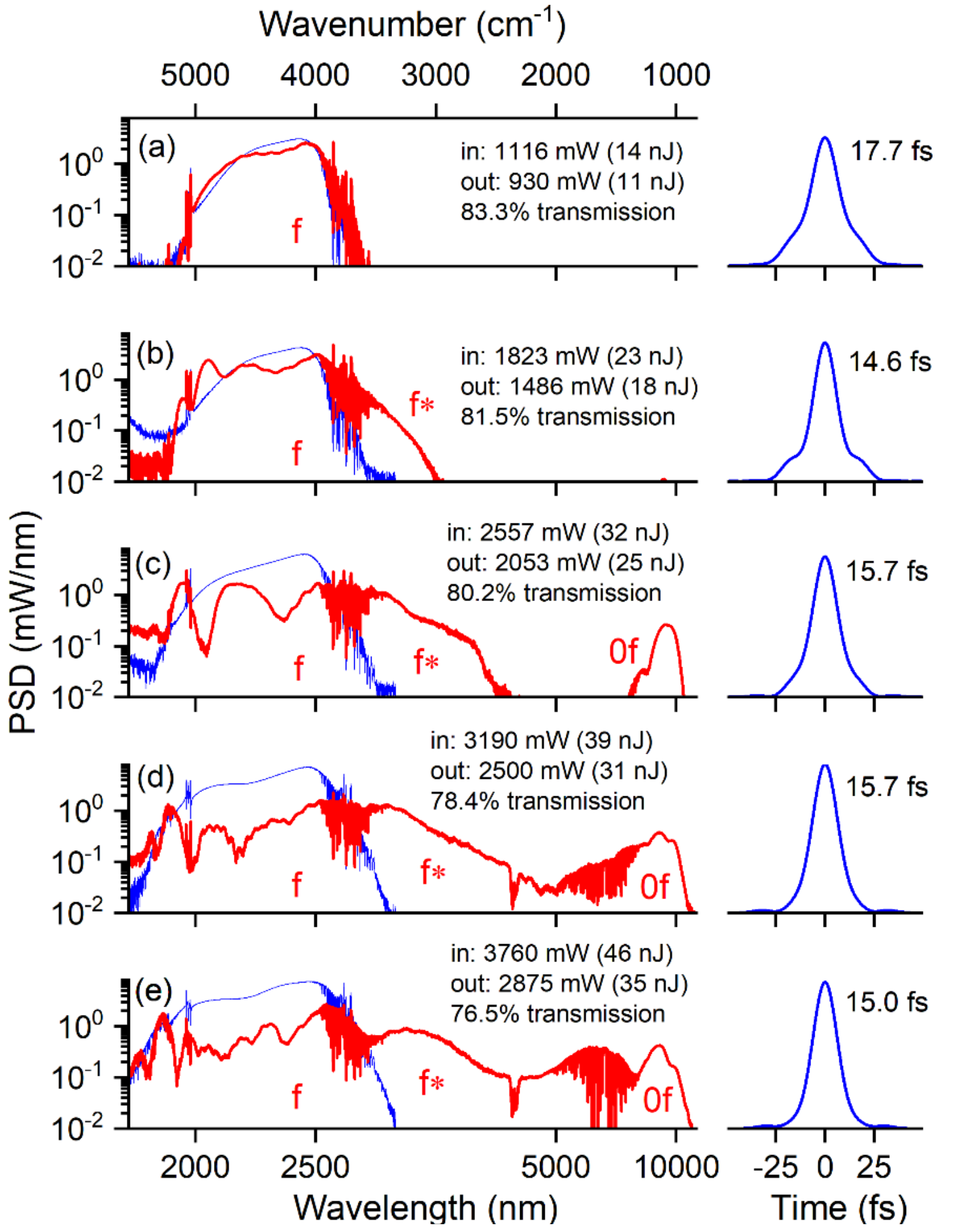

Figure S2. Output spectra of the mid-IR supercontinuum source as compared to the input spectra from the Cr:ZnS comb source (thin blue lines). Vertical axes, power spectral density (PSD) in log scale; bottom horizontal axis, wavelength in reciprocal scale; top vertical axis, wavenumber; f and f*, fundamental, red-shifted fundamental spectral bands; 0f, offset-free spectral band. Right panels show temporal intensity distributions of input pulses derived from the spectra assuming a flat spectral phase. Numbers show the pulse width at half maximum. The spectra were measured at the axis of the laser beam with a Thorlabs OSA207C Fourier transform optical spectrum analyzer (spectral range 1.0 – 12 µm) and then normalized to the measured average power. Therefore the power spectral density (PSD) distributions are approximate.

## 2. The linewidth of a free-running Cr:ZnS comb source

The linewidth of a free-running Cr:ZnS comb source was evaluated by the analysis of the heterodyne beating $f_B$ between a spectral component of the free-running comb and a Menlo Systems ORS-mini reference laser at the wavelength 1542.14 nm and with the linewidth 0.5 Hz (measured in a 2.6 s window) [5]. The setup for the generation of the beating signal $f_B$ is illustrated in Figure S3. A near-IR part of the comb's spectrum is separated from the main mid-IR signal with a dichroic mirror (DM). A residual SF pump radiation (up to 3 W at the wavelength 1567 nm) is separated with an Optigrate notch filter (NF, optical density 7). The optical signal is further bandpass filtered (BP) to isolate the frequency comb components in the vicinity of the wavelength 1542 nm. The bandpass filtered optical signal is superimposed with the reference laser on a broadband 50:50 fiber optic coupler, and then the beat signal $f_B$ is detected with a balanced InGaAs photodetector. The $f_B$ signal near the 17 MHz RF central frequency was properly low-pass filtered, amplified, and digitized at the 250 MHz sampling rate. A spectrogram of the digitized signal is illustrated in Figure S4 (a).

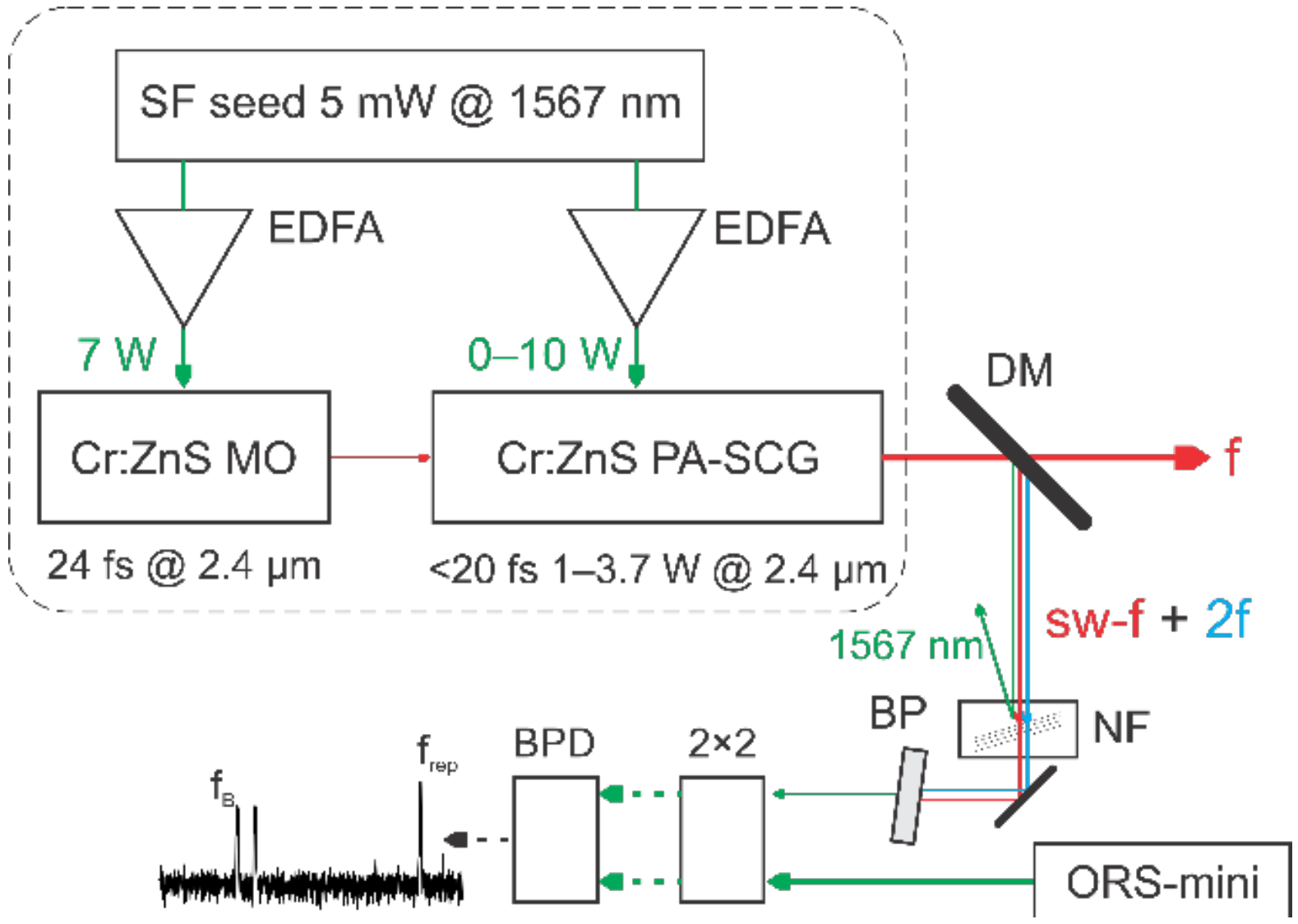

Figure S3. Dashed rectangle, Cr:ZnS comb source (see text). 1567, residual pump radiation; sw-f, 2f, near-IR components of the fundamental comb and its second harmonic; NF, notch filter; BP, bandpass filter, 2×2, fiber optic coupler, BPD, balanced photodetector.

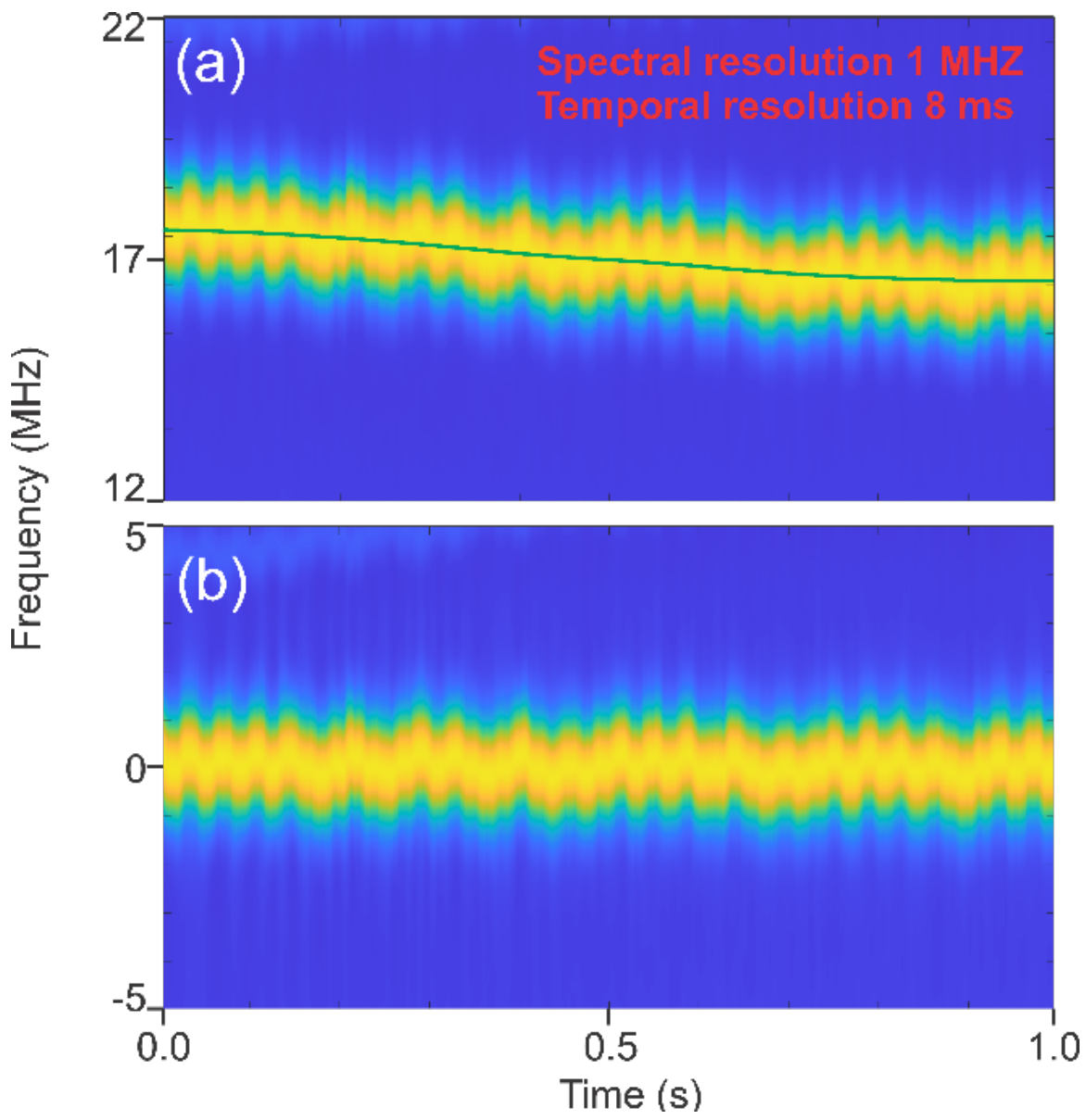

Figure S4. (a) Spectrogram of digitized $f_B$ signal measured during 1 s near 17 MHz central frequency in 5 MHz spectral span; green line shows slow trend of the signal's instantaneous frequency. (b) Spectrogram of de-trended $f_B$ signal.

The obtained data were post-processed to remove a slow drift from the beat signal, as shown in Figure S4 (b). We then studied the de-trended $f_B$ signal to identify its 'quiet' parts that are relatively free from noise, e.g., an acoustical noise. The linewidth of the comb was estimated by computation of the power spectra of the $f_B$ signal on different time scales. The obtained results are summarized in Figure S5. The linewidth of the beat signal reached the minimum of 5.6 kHz in a 0.5 ms window in the most 'quiet' part of the signal, see Fig. S5(a). The power spectra of the beat signal in the vicinity of this minimum are illustrated in Fig. S5(a). Additional analysis has shown that the linewidth reaches the minimum of about 12.5 kHz during 0.2 ms in other quiet parts of the signal.

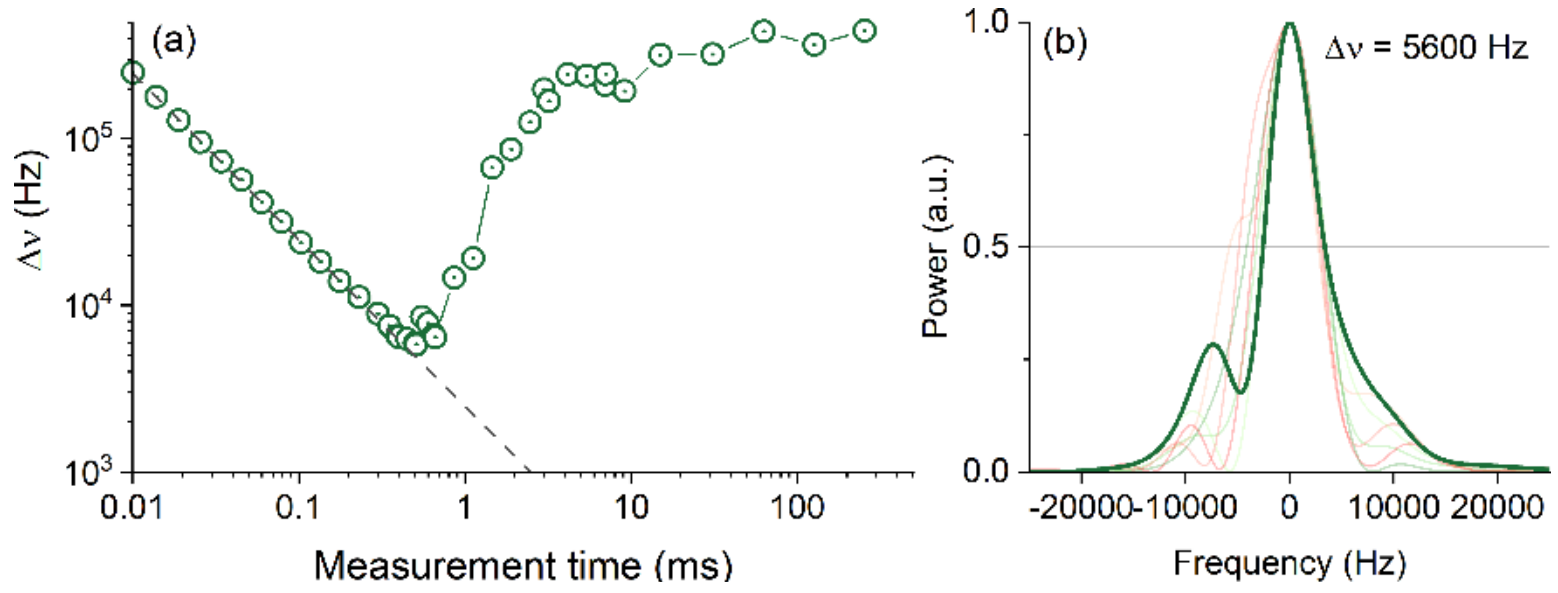

Figure S5. (a) FWHM of the power spectrum of the de-trended $f_B$ signal vs. measurement time; the dashed line shows the resolution bandwidth limit. (b) Power spectra of $f_B$ signal in the vicinity of the minimum. The thick line corresponds to a 0.5 ms measurement time. Thin lines correspond to 0.4, 0.45, 0.55, 0.6 ms measurement times.

Thus, the obtained estimate for the linewidth of the free-running 2.4-µm frequency comb source (~ 10 kHz) shows that polycrystalline oscillators feature not only intrinsically low intensity noise, but also low phase noise.